\documentclass[dvipdfmx,twocolumn,superscriptaddress,showpacs,preprintnumbers,amsmath,amssymb,pra]{revtex4}
\usepackage[dvipdfmx]{graphicx}
\usepackage{textcomp}   
\usepackage[scaled]{helvet}  
\usepackage{amsmath,amssymb}  
\usepackage{type1cm}
\usepackage{bm}  
\usepackage{color}
\usepackage{ulem}
\usepackage{comment}

\begin{document}

\title{Entanglement propagation in thermalization of an isolated quantum system}

\author{Ryosuke Yoshii} 
\affiliation{Center for Liberal Arts and Sciences, Sanyo-Onoda City University, 1-1-1 Daigaku-Dori, Sanyo-Onoda, Yamaguchi 756-0884, Japan}

\author{Shion Yamashika}
\affiliation{Department of Physics, Chuo University, 1-13-27 Kasuga, Bunkyo-ku, Tokyo 112-8551, Japan}

\author{Shunji Tsuchiya}
\affiliation{Department of Physics, Chuo University, 1-13-27 Kasuga, Bunkyo-ku, Tokyo 112-8551, Japan}

\date{\today}
\begin{abstract}
We study dynamics of entanglement in the thermalization process of an isolated quantum many-body system. 
We propose a simple setup for measuring the propagation speed of entanglement entropy (EE) in numerical simulations 
and apply it to the integrable/non-integrable spin models in 1D - the transverse Ising (TI) model, the chaotic Ising (CI) model, and the extended chaotic Ising (ECI) model.
We find that two distinct time-scales $t^\ast$ and $t_{\rm {diff}}$ arise in the dynamics of EE in the thermalization process: 
the former represents the time-scale for the saturation of EE and the latter characterizes spreading of EE over the entire system. 
Evaluating the propagation speed of entanglement from $t_{\rm diff}$, 
we find that entanglement propagates ballistically with a constant velocity irrespective of the integrability of the model. 
The propagation speed of entanglement is found to coincide with the maximum group velocity of quasi-particle excitations in the TI model. 
We also evaluate the propagation speed of entanglement by mutual information and find the characteristic time-scale $t_{\mathrm{MI}}$. 
We show that the propagation speeds of entanglement evaluated by $t_{\mathrm{MI}}$ and $t_{\rm {diff}}$ agree well. 
We discuss the condition for thermalization based on the numerical results 
and propose that scrambling of the entire system has to take place before saturation of EE for thermalization.

\end{abstract}

\maketitle

\section{Introduction}

Entanglement is a key concept in modern quantum theory and quantum information science \cite{nielsen10}. 
The importance of the notion of entanglement was first recognized by Einstein, Podolsky, and Rosen (EPR) in their attempt to demonstrate 
that quantum theory does not provide a complete description of physical reality, 
where a pair of particles in an entangled state was employed in the thought experiment they proposed \cite{EPR}. 
Motivated by the work of EPR, Bell deduced inequalities that should be satisfied by local hidden variable theories \cite{Bell,CHSH}. 
He revealed the striking feature of an entangled state: It possesses non-local correlations and violates those inequalities. 
It suggests that quantum theory violates local realism. 
The experimental verifications of the violation of the Bell inequalities pioneered by Aspect et al.\,\cite{Aspect1,Aspect2,Aspect3} implies 
that local realism is fundamentally violated in nature, though there remains possibility of loop holes in the experiments.

Recently, entanglement attracts renewed interests in the study of quantum many-body systems \cite{amico08}. 
It has been realized that entanglement is a key to understand novel aspects of quantum many-body systems 
such as quantum phase transitions and topological phases \cite{kitaev03,zeng19}. 
One of the quantities that characterizes entanglement structure of a quantum many-body system is entanglement entropy (EE). 
Although EE has been intensively studied theoretically in diverse fields \cite{amico08,nishioka18}, measuring EE in experiments remains challenging in usual solid state systems.
Cold atoms have emerged as an ideal platform for studying entanglement in many-body systems due to the unprecedented controllability. 
It is remarkable that the second-order R\'enyi entropy, which also quantifies entanglement, 
for atoms in an optical lattice has been successfully measured making use of a quantum gas microscope \cite{islam15}.

Progress of cold atom experiments has also enabled investigations of fundamental issues in statistical mechanics \cite{abanin19}. 
It is especially suitable for studying the mechanism of thermalization in an isolated quantum system \cite{deutsch91,srednicki94,tasaki98,ikeda15,kinoshita06,rigol08}.
Experimental studies have elucidated that thermalization under the unitary evolution of the whole system occurs on local scale 
due to the growth of EE in subsystems \cite{kaufman16}.

Propagation speed of entanglement in a quantum many-body system has also been a fundamental issue and addressed theoretically \cite{amico08,lieb72,calabrese05,kim13,bravyi06,kormos17}. 
Propagation of correlations have been experimentally studied using cold atoms \cite{cheneau12} as well as trapped ions \cite{richerme14}. 
It corresponds to the speed of light in relativistic theories and restricts propagation of information. 
It is deeply related with thermalization in a quantum system and plays a crucial role in the proof of the second law of thermodynamics and the nonequilibrium fluctuation theorem \cite{iyoda17,iyoda18}. 

Since entanglement is not a conserved quantity that can be locally created, a careful analysis is required to study its transport.
In Ref.~\cite{calabrese05}, for example, propagation speed of entanglement, 
which is identified as that of quasi-particles, is extracted from the time-evolution of EE in a subsystem after a quench assuming that the characteristic time for saturation of EE, 
referred to as $t^\ast$, is related with the maximum propagation speed of entanglement $V$ as $t^\ast=L/2V$, where $L$ is the subsystem size. 
However, the interpretation of the saturation of EE based on the quasi-particle picture is not valid for systems without well-defined quasi-particles. 
Thus it is not clear if the propagation speed of EE manifests itself in the subsystem size dependence of $t^\ast$ in such systems. 
Later the time evolution of EE after a global quench in a strongly coupled gapless system has been analyzed by the holographic calculations in Ref.\,\cite{Liu14}. 
The linear growth of EE has been found and has been interpreted due to the spread of the wave front of the ``Tsunami" of entanglement. 
The idea of the ``entanglement Tsunami" has been further examined by the 1+1 dimensional CFT in the limit of large central charge 
and it has been shown that the time dependence of EE in multiple intervals is not correctly captured by the quasi-particle model \cite{Leichenauer15}.
It has also been shown that this entanglement Tsunami picture is no longer valid for the small subsystems \cite{Kundu17}.
Thus it is expected that a further investigation of the spread of entanglement in finite systems without free quasi-particle picture yields deep insights into entanglement transport.

In this paper, we study dynamics of entanglement in an isolated quantum many-body system. 
In order to detect entanglement propagation, we propose a control experiment 
in which the propagation of locally created entanglement is detectable by monitoring the difference of EE for two systems with different initial settings.
We demonstrate it in numerical simulations for the transverse Ising (TI) model, the chaotic Ising (CI) model, and the extended chaotic Ising (ECI) model. 
We also study entanglement propagation by analyzing mutual information.
Here we summarize the main findings of the paper. 
\begin{itemize}
\item{}
Two time scales arise in the dynamics of EE: $t^\ast$ characterizing the saturation of EE and $t_{\rm diff}$ characterizing the spreading of entanglement. 
\item{}
Two time scales $t^\ast$ and $t_{\rm diff}$ do not coincide. The propagation speed of EE can be evaluated by the latter. 
\item{}
The propagation speed of entanglement for the TI model evaluated by $t_{\rm diff}$ is found to coincide with the maximum group velocity of quasi-particle excitations. 
\item{}
The propagation of entanglement is also detectable by measuring mutual information. 
The propagation speeds of entanglement evaluated by mutual information and EE agree well. 
\item{}
The time scale $t^\ast$ is larger than $t_{\rm diff}$ for the ECI model which achieves the thermalization. 
$t^\ast$ is smaller than $t_{\rm diff}$ for the TI and CI models which do not achieve thermalization. 
\end{itemize}
Based on the above results, we further discuss the conditions for thermalization based on the numerical results 
and we propose that scrambling of the entire system has to take place before saturation of EE for thermalization.

This paper is organized as follows: 
We introduce the model in Sec.\,\ref{model} and study the dynamics of total magnetization and EE in the thermalization process in Sec.~\,\ref{dynamicsofEE}. 
In Sec.\,\ref{SetupNS}, we describe the setup for numerical simulations. 
We apply it to the spin models to study the propagation speed of EE in Sec.\,\ref{PropEE}. 
We study propagation of entanglement by calculating mutual information in Sec.\,\ref{MI}. 
We conclude and give a brief discussion about the condition for thermalization and experimental realization in Sec.\,\ref{Disc}. 

\section{model}
\label{model}

We consider a one dimensional spin chain that consists of $N$ spin-1/2s described by the Hamiltonian 
\begin{equation}
H=-J\sum^{N-1}_{i=1} \sigma^z_i\sigma^z_{i+1}-J^\prime \sum^{N-2}_{i=1} \sigma^z_i\sigma^z_{i+2}
+\sum_{i=1}^N\left(h_z \sigma^z_i+h_x \sigma^x_i\right),
\label{Hamiltonian}
\end{equation}
where $\sigma_i^\mu$ ($\mu=x,y,z$) are the Pauli matrices at the $i$-th site. 
We set $\hbar=1$, $J=1$, and assume zero temperature throughout the paper. 
Specifically, we study three models in this paper: 
The TI model, the CI model, and the ECI model. 
We set the parameters $J^\prime=0$, $h_z=0$ for the TI model, $J^\prime=0$, $h_x=1.05$, $h_z=0.5$ for the CI model, and $J^\prime=0.8$, $h_x=1.05$, $h_z=0.5$ for the ECI model. 
We denote the spin-up and spin-down states at the $i$-th site as $|\uparrow\rangle_i=|0\rangle_i$ and $|\downarrow\rangle_i=|1\rangle_i$, respectively.
In our numerical simulations, we calculate eigenvalues and eigenvectors of the Hamiltonian (\ref{Hamiltonian}) by exact diagonalization.

\section{dynamics of entanglement entropy in thermalization}
\label{dynamicsofEE}

We first discuss dynamics of EE in thermalization process \cite{rigol08}. 
In an isolated system, the wave function for the whole system evolves in time as $|\psi(t)\rangle=\sum_n a_n e^{-iE_n t}|n\rangle$, 
where $E_n$ is the energy eigenvalue for the energy eigenstate $|n\rangle$ and $a_n=\langle n|\psi(0)\rangle$. 
The average of an observable $A$ is given by
\begin{equation}
\langle A(t)\rangle=\langle\psi(t)|A|\psi(t)\rangle=\sum_{n,m}a_n^*a_me^{i(E_n-E_m)t}\langle n|A|m\rangle.
\label{diagonalave}
\end{equation}
In the relaxation to equilibrium, the off-diagonal terms in Eq.~(\ref{diagonalave}) vanish due to interference and the long-time average 
thus approaches the diagonal average $\langle A\rangle_{\rm diagonal}=\sum_{n}|a_n|^2 \langle n|A| n\rangle$, if the eigenvalues $E_n$ are not degenerate. 
The system is considered to be thermalized if the diagonal average is equivalent to the microcanonical average
\begin{equation}
\langle A\rangle_{\rm microcan}(E_0)=\frac{1}{\mathcal N_{E_0,\Delta E}}\sum_{|E_n-E_0|<\Delta E}\langle n|A| n\rangle.
\label{microcanonical}
\end{equation} 
Here, $E_0=\langle\psi(0)|H|\psi(0)\rangle$ is the total energy of the initial state 
and the summation in Eq.~(\ref{microcanonical}) is taken over the states in the energy shell that has the width $2\Delta E$ around $E_0$. 
$\mathcal N_{E_0,\Delta E}$ is the number of states in the energy shell and we set $\Delta E=0.2$ in the following numerical calculations.
\par
We calculate time-evolution of total magnetization $M_z=\sum_{i=1}^N\langle\sigma_i^z\rangle$. 
We set the system size $N=14$ and choose the N\'eel state $|\psi(0)\rangle=|0\rangle_1|1\rangle_2|0\rangle_3|1\rangle_4\dots\equiv|\text{N\'eel}\rangle$ as the initial state.
Whether the system thermalizes or not depends on the integrability/non-integrability of the model. 
It is well known that the TI model is integrable because it can be diagonalized by the Jordan-Wigner transformation \cite{sachdevbook}, whereas the CI and ECI models are non-integrable \cite{iyoda18,banuls11}.
\par
\begin{figure}[t]
\includegraphics[width=7.5cm]{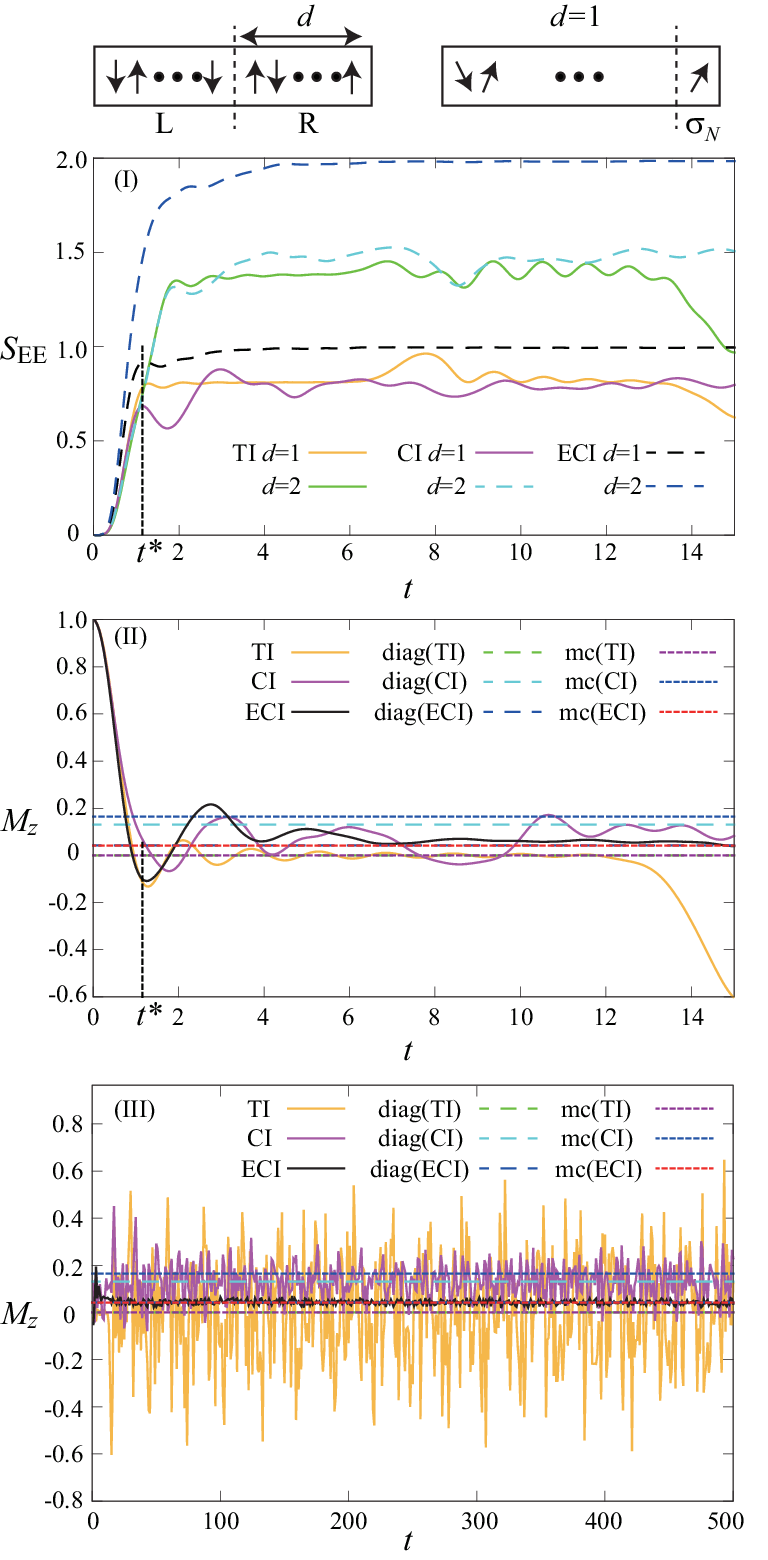}
\caption{Schematic picture of spin chain (top figure), (I) Time-evolution of EE for the subsystem R, and (II) that of local magnetization $M_z=\langle\sigma_N^z\rangle$ for the TI, CI, and ECI models. 
The diagonal and microcanonical averages for each model are, respectively, indicated as the broken horizontal line and the dotted horizontal line in (II). 
The diagonal and microcanonical average are, respectively, $0.131$ and $ 0.165$ for the CI model,  
$-1.03\times 10^{-15}$ and $ -1.86\times 10^{-17}$ for the TI model, and $4.24\times 10^{-2}$ and $ 4.16\times 10^{-2}$ for the ECI model.  
Fig.\ (III) shows the longtime behaviour of local magnetization $M_z$.} 
\label{figSzTMLZ}
\end{figure}

Figure~\ref{figSzTMLZ} (II) demonstrates thermalization in the ECI model as the magnetization relaxes to the diagonal average that is close to the microcanonical average. 
On the other hand, thermalization is not achieved in the TI model as well as the CI model: 
The former shows a recurrence behavior of magnetization, 
while in the latter magnetization exhibits large fluctuation around the diagonal average that does not coincide with the microcanonical average 
(see Fig.\ \ref{figSzTMLZ} (II)). 
Here, we note that whether the CI model thermalizes or not for sufficiently large system size is still controversial. In fact, one of the previous studies claims that thermalization is not achieved in the CI model \cite{banuls11}.
\par
Throughout this paper, we focus on the system with $N=14$. We confirm that the finite size effect is irrelevant for the following analysis (See Appendix A). 

We compare time evolution of EE and that of magnetization. 
If the whole system is divided into the subsystems L and R (see Fig.~\ref{figSzTMLZ} (II)), EE for the subsystem R can be calculated as \cite{amico08}
\begin{equation}
S_{\rm R}=-\mathrm{Tr_ R}(\rho_{\rm R}\log\rho_{\rm R}).
\end{equation}
Here, $\rho(t)=|\psi(t)\rangle\langle\psi(t)|$ and $\rho_{\rm R}(t)=\mathrm{Tr}_{\rm L}\rho(t)$ are the density matrices for the whole system and the subsystem R, respectively. 
$\rm Tr_R$ ($\rm Tr_L$) stands for taking trace over the subsystem R (L). 
In the following, since we only consider pure states, $S_{\rm R}$ coincides with $S_{\rm L}=-\mathrm{Tr}_L(\rho_L\log\rho_L)$, where $\rho_L=\mathrm{Tr}_R\rho(t)$ \cite{nielsen10}. 
Thus we denote EE as $S_{\rm EE}(=S_{\rm R}=S_{\rm L}).$

Figure~\ref{figSzTMLZ} (II) shows the time evolution of EE for the subsystem R. 
EE for the ECI model grows linearly in the initial stage and then after a characteristic time $t^\ast$ saturates at the equilibrium value that is proportional to the size of the subsystem $d$, 
which is known as the volume law \cite{amico08,abanin19,calabrese05,kaufman16}. 
The saturation of EE implies thermalization of the system in the time scale of $t^\ast$. 
The TI model and CI model exhibit similar behaviors, for which $t^\ast$ can be defined as in Fig.~\ref{figSzTMLZ} (I). 
The large fluctuation after the saturation, however, implies that the system does not thermalize, which is consistent with the behavior of magnetization. 
By comparing the TI model and the CI model, it is clear that the presence of the additional $h_z$ term in the CI model highly suppresses the fluctuation.

Comparing Figs.~\ref{figSzTMLZ} (I) and (II), both EE and magnetization saturate at the equilibrium values in the same time scale $t^\ast$, 
which can be qualitatively understood as follows \cite{calabrese18}.
Local magnetization can be written as 
\begin{equation}
M_z=\langle \sigma^z_N\rangle=\mathrm{Tr}(\sigma_N^z\rho(t))=\mathrm{Tr}_N(\sigma_N^z\rho_N(t)),
\label{lmag}
\end{equation}
where $\rho_N=(\prod_{j\neq N}\mathrm{Tr}_j)\rho$ is the reduced density matrix for site $N$. 
It turns out from Eq.~(\ref{lmag}) that time-evolution of local magnetization reflects that of the single-site density matrix $\rho_N(t)$.
The thermal equilibration in $\rho_N(t)$ thus leads to the saturation of both EE and local magnetization for a single site in the same time scale of $t^\ast$ 
(see the right figure of the schematic picture on the top of Fig.\ \ref{figSzTMLZ}). 

\section{Setup for numerical simulations}
\label{SetupNS}

In time-evolution under the Hamiltonian~(\ref{Hamiltonian}), entanglement between neighboring spins is generated locally by the spin-spin interaction terms. 
Meanwhile, they create correlations between distant spins and entanglement propagates in time as the number of correlated spins grows. 
EE for the subsystem R increases linearly in the early stage of the time-evolution, as shown in Fig.~\ref{figSzTMLZ} (I), 
not only because entanglement generated in the subsystem L propagates and spreads over the subsystem R, but also entanglement is locally generated in the vicinity of the subsystem R. 
Therefore, both the time scales of the entanglement propagation and the local entanglement generation should be encoded in the characteristic time $t^\ast$. 
However, it is not clear how $t^\ast$ is related with the two time scales.
\par
\begin{figure}[t]
\includegraphics[width=7.5cm]{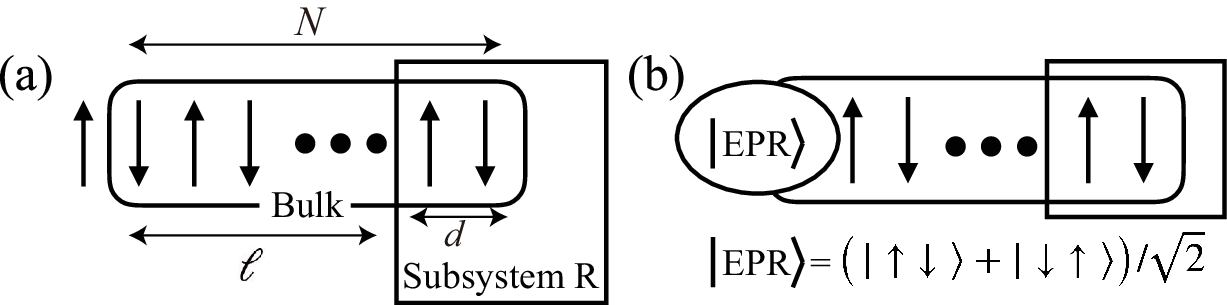}
\caption{Setup for measuring propagation speed of entanglement. A single spin is attached to the spin chain (bulk part). 
The attached single spin and the bulk part are in a product state in the initial state (a). 
The attached spin and a single spin on the left end of the bulk part are entangled in the initial state (b). 
EE is evaluated for the subsystem R on the right end of the bulk part with the size $d$. $\ell =N-d$ denotes the propagation distance of EE from the left end to the subsystem R.}
\label{figmodel}
\end{figure}

To study propagation speed of entanglement, it is necessary to evaluate time for entanglement to propagate a certain distance. 
For this purpose, we make a comparison of the time-evolution of EE for the two cases that start from the different initial states, schematically shown in Fig.~\ref{figmodel}. 
We consider a system that consists of $N+1$ spins where a single spin at the site $i=0$ is attached on the left end of the spin chain on the sites $1\leq i\leq N$. 
We hereafter refer to the spin chain without the attached spin as the ``bulk part".
In studying dynamics, we let the bulk part evolve in time under the Hamiltonian (\ref{Hamiltonian}).

In the initial state (a), the attached spin and the bulk part are in a product state as
 \begin{equation}
|\psi\rangle_a=|0\rangle_0\otimes |\Psi\rangle_{\rm bulk}.
\end{equation}
In the initial state (b), on the other hand, the two spins on the left end are maximally entangled as
\begin{equation}
|\psi\rangle_b=\Lambda_{0,1}\left\{\frac{|0\rangle_0+|1\rangle_0}{\sqrt{2}}\otimes |\Psi\rangle_{\rm bulk}\right\},
\label{entini}
\end{equation}
where $\Lambda_{0,1}$ denotes the CNOT gate \cite{nielsen10} with the site $i=0$ being the control gate and the site $i=1$ the target gate. 
We note that the initial state (b) was introduced in Ref.~\cite{iyoda18}. 
In the following numerical simulations, the initial state of the bulk part $|\Psi\rangle_{\rm bulk}$ is set to the N\'eel state $|\Psi\rangle_{\rm bulk}=|\text{N\'eel}\rangle$ for simplicity. 
The initial state (b) thus has an EPR pair on the left end.
We performed the same numerical simulations described below taking other product states including a ferromagnetic state as $|\Psi\rangle_{\rm bulk}$ 
and confirmed the following arguments do not depend on the specific choice of $|\Psi\rangle_{\rm bulk}$. 

\section{Propagation of entanglement entropy}
\label{PropEE}

We compare time-evolution of EE for the three models in order to study propagation of EE in detail.
We evaluate EE for the subsystem R with the length $d$ on the right end of the bulk part as shown in Fig.~\ref{figmodel}. 
$\ell=N-d$ denotes the distance for entanglement to travel from the left end to the subsystems. 
We calculate EEs for each model with $N=14$.
It is expected that a comparison between these three models reveals the role of entanglement for thermalization.

\subsection{Chaotic Ising model}
\begin{figure}[t]
\includegraphics[width=7.5cm]{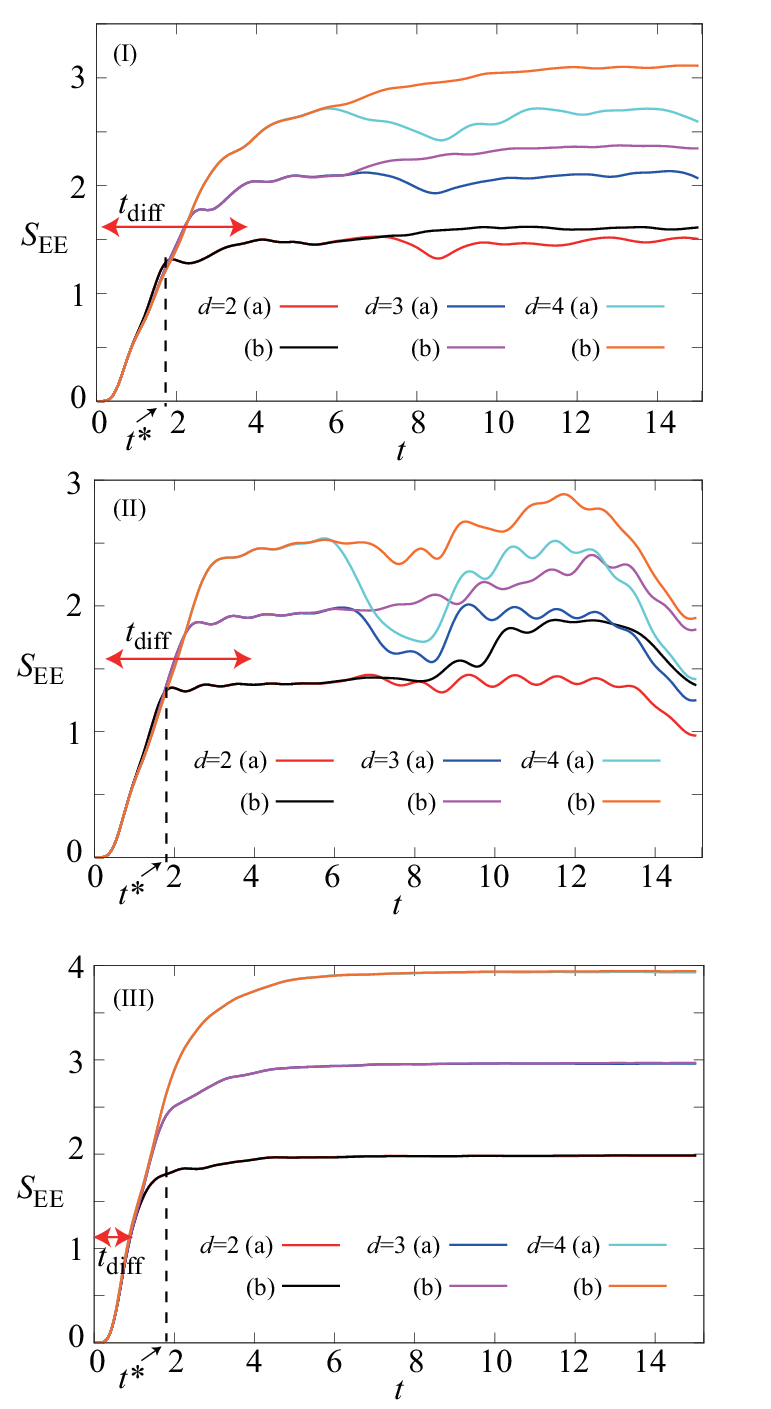}
\caption{Comparison of time-evolution of EE starting from the initial states (a) and (b) for the (I) CI model, (II) TI model, and (III) ECI model. 
Arrows indicate $t_{\rm MI}$ for $d=2$ ($\ell=8$). 
The time scale $t^\ast$ is indicated by the dashed vertical lines in each figure. 
$t^\ast$s are chosen to be the same for three models due to the difficulty of the quantitative definition. } 
\label{figvelEEs}
\end{figure}

Figure~\ref{figvelEEs} (I) shows the time-evolution of EE for the CI model. 
The curves for the initial states (a) and (b) precisely coincide during the linear increase in $t<t^\ast$ and they split at $t_{\mathrm{diff}}(>t^\ast)$ after saturation. 
Here we set the threshold to be $10^{-7}$, namely the time $t_{\mathrm{diff}}$ is defined by the shortest time at which the difference between the EE for different initial states exceeds $10^{-7}$. 
We confirmed that the qualitative behaviour of $t_{\mathrm{diff}}$ does not change by changing the threshold in the range of $10^{-6}\sim 10^{-9}$. 
\par 
The new characteristic time $t_{\mathrm{diff}}$ can be interpreted as the moment at which entanglement on the left end in the initial state (b) reaches the subsystem R. 
It means that entanglement spreads over the system in the time-scale of $t_{\mathrm{diff}}$. 
This interpretation is consistent with the fact that $t_{\rm diff}$ increases as the travel distance of entanglement $\ell$ increases as shown in Fig.~\ref{figvelEEs} (I). 
\par
An advantage of considering $t_{\mathrm{diff}}$ is that 
we can calculate the propagation speed of entanglement without considering $d$-dependence of $t_{\rm diff}$ for a fixed large $\ell$.
\par
It is remarkable that the saturation and the spreading of entanglement occur in the two different time-scales $t^\ast$ and $t_{\rm diff}$, respectively.
In view of the fact that splitting of the curves for (a) and (b) takes place after saturation, $t^\ast$ may be related to the time-scale for local generation of entanglement, 
where the rate of local generation of entanglement determines the saturation time $t^\ast$. 
We discuss about $t^\ast$ in more detail in Sec.~VII.

Figure~\ref{figtransvel} (II) shows $\ell=N-d$ as a function of $t_{\rm diff}$ for the CI model. 
The propagation speed of entanglement can be evaluated by the slope of the curve. 
The constant slope in Fig.~\ref{figtransvel} (II) indicates ballistic propagation of entanglement with a constant velocity $v_{\rm CI}\simeq 2$. 
We confirmed that the propagation speed evaluated in this manner depends on neither the specific choice of the initial state nor the system size. 
Although the same conclusion has been obtained in Refs.~\cite{kim13,iyoda18}, the analyses in these papers are based on the evaluation of $t^\ast$ that may not be directly related to entanglement propagation.

\subsection{Transverse Ising model}

\begin{figure}[t]
\includegraphics[width=7.5cm]{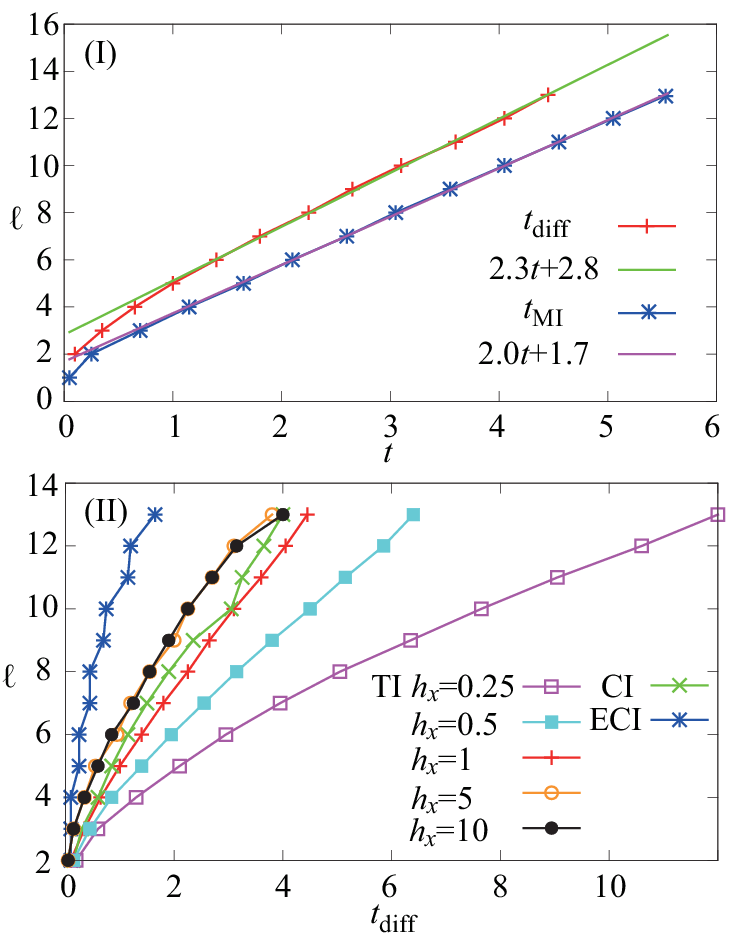}
\caption{(I) $t_{\mathrm{diff}}$ and $t_{\rm MI}$ are plotted as a function of $\ell$ for the TI model with $h_x=1.0$. 
The propagation speeds of EE and mutual information are evaluated from the slope of the curves as $v_{\rm diff}=2.3$ and $v_{\rm MI}= 2.0$. 
(II) $t_{\mathrm{diff}}$ is plotted as a function of $\ell$ for the TI model for various strength of the transverse field $h_x$. 
Propagation speed of entanglement $v_{\rm TI}$ is proportional to $h_x$ as $v_{\rm TI}\simeq 2h_x$ for $h_x<1$, while it is a constant $v_{\rm TI}\simeq 2.0$ for $h_x\ge 1$. 
} 
\label{figtransvel}
\end{figure}

We apply the same analysis made above to the TI model in order to examine what determines the propagation speed of entanglement. 
Figure \ref{figvelEEs} (II) shows the time-evolution of EE for the TI model. 
The similar behaviour to the CI model is found. 
Namely, the curves for the initial states (a) and (b) precisely coincide during the linear increase in $t<t^\ast$ and they split at $t_{\mathrm{diff}}(>t^\ast)$ after saturation. 
Figure \ref{figtransvel} (II) shows $\ell$-$t_{\rm diff}$ plot for various strength of the transverse magnetic field. 
It demonstrates that entanglement spreads with a constant speed, because each curve has a constant slope.
The slope of the curve increases monotonically as $h_x$ increases for $h_x< 1$, whereas it is almost independent of $h_x$ for $h_x\ge 1$. 
Thus, the propagation speed of entanglement $v_{\rm TI}$ can be evaluated as
\begin{eqnarray}
v_{\rm TI}\simeq \left\{
\begin{array}{ll}
2h_x,\quad &(h_x<1),\\
2,\quad &(h_x\ge 1).
\end{array}
\right.
\label{eqvti}
\end{eqnarray}

The TI model can be diagonalized by the Jordan-Wigner transformation \cite{sachdevbook} as
\begin{equation}
H=\sum_k\varepsilon_k\left(\gamma_k^\dagger \gamma_k-\frac{1}{2}\right),
\end{equation} 
where $\gamma_k^\dagger$ is the creation operator of a quasi-particle with momentum $k$. The energy dispersion of quasi-particles is given by
\begin{equation}
\varepsilon_k=2\sqrt{1+h_x^2-2h_x\cos k}.
\end{equation} 
The maximum group velocity of quasi-particles is found to be
\begin{eqnarray}
v_{\mathrm{max}}=\mathrm{max}_k\{\partial \varepsilon_k/\partial k\}=\left\{
\begin{array}{ll}
2h_x,\quad &(h_x<1),\\
2,\quad &(h_x\ge 1).
\end{array}
\right.
\end{eqnarray}
The same dependence of $v_{\rm TI}$ and $v_{\mathrm{max}}$ on $h_x$ thus indicates that EE propagates with quasi-particle excitations. 
We note that $v_{\rm TI}$ in Eq.~(\ref{eqvti}) is much below the upper bound $12e$ determined by the Lieb-Robinson theorem applied to the TI model \cite{premont-schwarz10}. 
\par
Figure \ref{figtransvel} (II) shows that the curve for the CI model has the same slope as those for the TI model with $h_x\ge 1$. 
It suggests that the longitudinal magnetic field does not affect propagation speed of entanglement 
and strongly supports the validity of the quasi-particle picture for transport of entanglement in the CI model despite its non-integrability.

\subsection{Extended chaotic Ising model}

Figure~\ref{figvelEEs} (III) shows time-evolution of EE for the ECI model. The saturation of EE clearly demonstrates thermalization of the system.
The two time scales $t^\ast$ and $t_{\rm diff}$ also arise analogous to the CI model. 
$t_{\mathrm{diff}}$ is shorter than those for the TI and CI models due to the next-nearest-neighbor (NNN) interaction, since it directly transports entanglement between the NNN sites. 
In contrast to the CI model, however, the two curves for the initial states (a) and (b) split at $t_{\rm diff}$ much earlier than the saturation at $t^\ast$, {\it i.e.}, $t_{\rm diff}\ll t^\ast$. 
It indicates that thermalization is achieved after entanglement spreads over the system.

The propagation speed of entanglement can be evaluated from the $\ell$-$t_{\rm diff}$ curve for the ECI model, 
which is not straight due to the even-odd dependence of $t_{\rm diff}$, in Fig.~\ref{figtransvel} (II). 
Since the NNN interaction directly couples the attached spin with the spins on the even sites, 
entanglement spreads on the even sites faster than the odd sites and thus $t_{\rm diff}$ is smaller for even $\ell$s than odd $\ell$s. 
The propagation speed, evaluated by the fitting in the range $1<t<6$ (the vicinities of edges are not included to avoid the finite size effect), 
$v_{\rm ECI}\simeq 5.5$ is much faster than $v_{\rm CI}$ and $v_{\rm TI}$ due to the NNN interaction.

\section{Mutual information}
\label{MI}

We demonstrate that propagation of entanglement can be also detected by measuring mutual information that quantifies bipartite correlation \cite{iyoda18, zeng19}. 
The mutual information between the attached spin and the subsystem R is given as $I(0|R)=S_0+S_R-S_{0R}$, where $S_0$, $S_R$, and $S_{0R}$ denote entanglement entropy for the attached spin, 
the subsystem R with length $d$, and their joint system, respectively. 
It is expected that $I(0|R)$ becomes finite when entanglement spreading from the left end gets to the subsystem R at $t_{\rm MI}$. 
Here we define $t_{\rm MI}$ to be the time at which $I(0|R)$ exceeds the threshold $|I(0|R)|>10^{-7}$.  
$t_{\rm MI}$ should coincide with $t_{\rm diff}$ since both characterize the propagation of entanglement traveling from the edge.  
It should be noted that the time scale $t^\ast$ is not detectable in the analysis of mutual information in the case of $t^\ast < t_{\rm MI} \simeq t_{\rm diff}$, 
since mutual information becomes finite after $t_{\rm MI}$ and thus the behavior of the mutual information before $t_{\rm MI}$ is invisible. 

\begin{figure}[t]
\includegraphics[width=7.5cm]{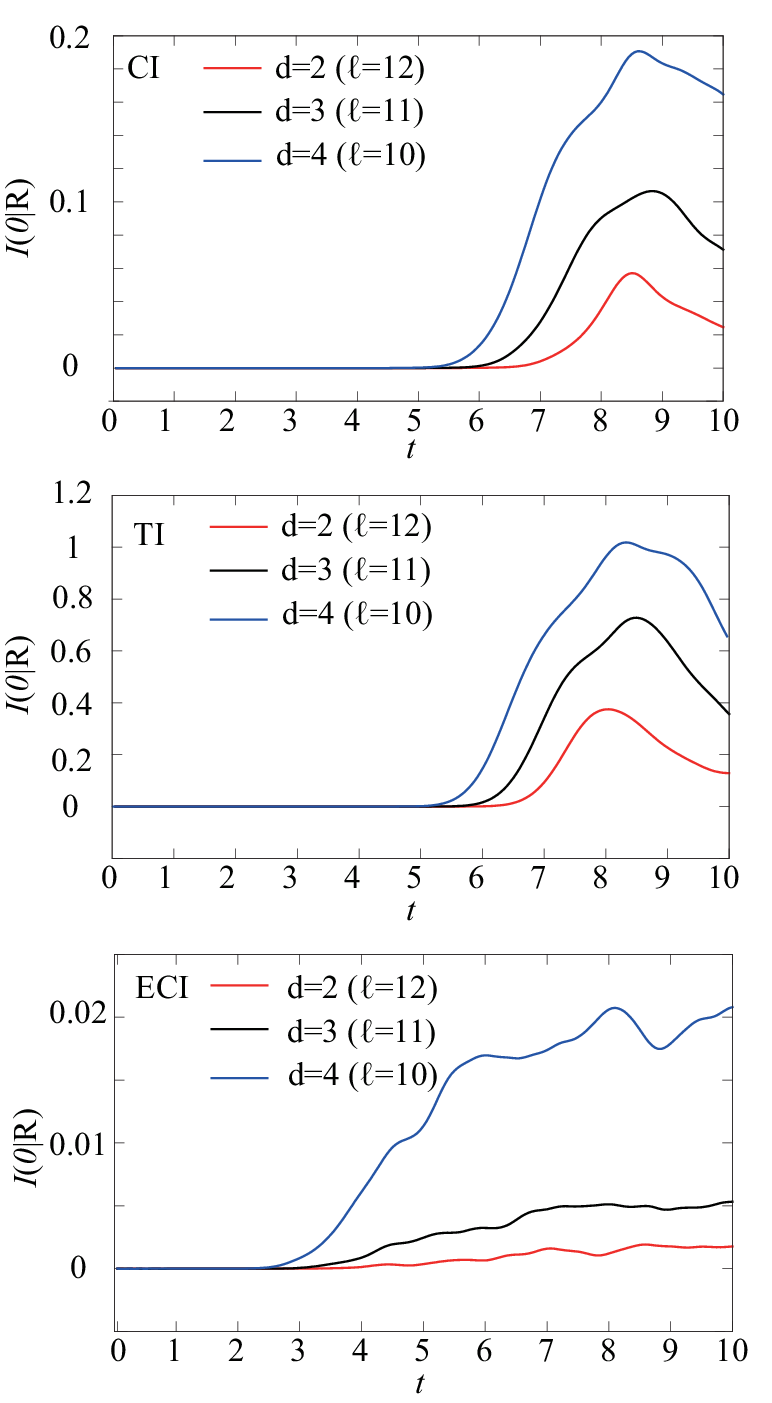}
\caption{Time dependence of $I(0|R)$ for CI, TI and ECI models.} 
\label{figMITD}
\end{figure}

Figure~\ref{figMITD} shows the time dependence of $I(0|R)$ for $d=2,3,4$ for each model. 
As expected, the mutual information becomes finite for $t>t_{\mathrm{MI}}$, whereas it is zero before $t_{\mathrm{MI}}$.

It should be noted that the magnitude of the mutual information significantly depends on the model which is in sharp contrast with the behaviour of EE; 
the saturation values of EE hardly depends on the model.

\begin{figure}[t]
\includegraphics[width=7.5cm]{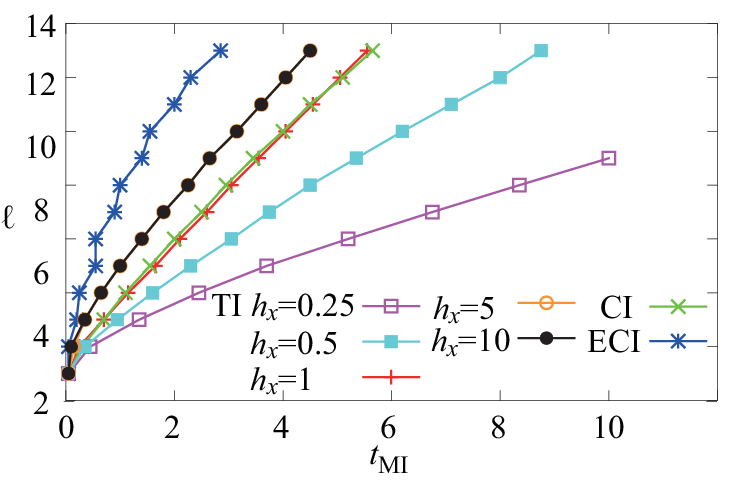}
\caption{Subsystem size dependence of $t_{\rm MI}$ for the TI model with $h_x=0.25,\,0.5,\,1.0,\,5.0,\,10 $, CI model, and ECI model. } 
\label{figChaosMIVel}
\end{figure}

Figure~\ref{figChaosMIVel} plots $t_{\rm MI}$ as a function of the subsystem size $\ell$ for the TI model with $h_x=0.25,\,0.5,\,1.0,\,5.0,\,10 $, CI model, and ECI model. 
It is found that the subsystem size dependence of $t_{\mathrm {MI}}$ shows the same behaviour as that of $t_{\mathrm {diff}}$ in Fig.\,\ref{figtransvel} (II).

\section{Conclusion and discussion}
\label{Disc}

We have examined the dynamics of entanglement in an isolated quantum system focusing on EE. 
In our numerical simulations, we found that the two time-scales arise in the thermalization process of an isolated quantum system: $t^\ast$ and $t_{\mathrm{diff}}$. 
$t^\ast$ represents the time-scale for saturation of EE.
Since the size of the Hilbert space for a subsystem with length $d$ is $2^d$, the maximum value that EE saturates is proportional to $d$ as ${\rm max}S_{\rm EE}\propto \log_2 2^{d}=d$, 
which explains the volume law. 
Since EE grows linearly in time before saturation reflecting the local generation of EE with a constant ratio, $dS_{\rm EE}/dt=\Gamma$ is a constant and thus we find $t^\ast\propto d/\Gamma$.
$t_{\rm diff}$, on the other hand, represents the time-scale for spreading of entanglement over the whole system.
In other words, $t_{\rm diff}$ characterizes how fast the system is scrambled.
The above interpretation is supported by the fact that $t^\ast$s hardly depend on the models as shown in Fig.\ \ref{figvelEEs}, 
whereas $t_{\mathrm{diff}}$s strongly depend on the models. 
It is natural to expect that the propagation speed of the entanglement for the ECI model is much faster than those for other two models due to the NNN coupling. 
Thus the results also suggest that the time sale characterizes the propagation of EE is not $t^\ast$ but $t_{\mathrm{diff}}$. 

Before moving to the discussion on thermalization, we make a few remarks. 
It is found that entanglement can be generated outside of the light cone and explained in terms of entanglement transfer by vacuum fluctuation \cite{Franson08, Reznik03, Reznik05, Koga18}. 
From these results on the acausal generation of entanglement, it should be natural to consider that the generation of entanglement is not only due to quasi-particles but also vacuum fluctuation. 

The numerical results $t_{\mathrm{diff}}\gtrsim t^\ast$ for the CI model and $t_{\mathrm{diff}}\ll t^\ast$ for the ECI model imply 
that scrambling of the entire system has to take place before saturation of EE for thermalization. 
The condition for thermalization can be derived from the following observation. Entanglement can propagate the distance $\xi=v_{\rm EE}\cdot t^{\ast}$ before EE saturates, 
where $v_{\rm EE}=\ell/t_{\rm diff}$ is the propagation speed of EE. 
If $\xi$ is much larger than the size of the subsystem $d$, \textit{i.e}. $\xi\gg d$, 
scrambling between the subsystem and other parts of the system occurs before local saturation of entanglement is achieved. 
As a result, the entire system is scrambled enough to realize thermal equilibrium when EE saturates. 
In the case of the ECI model, for example, $\xi\simeq 10$ for $d=2$ and thus $\xi\gg d$. 
If $\xi$ is smaller than $d$, \textit{i.e}. $\xi\lesssim d$ on the contrary, subsystems are not scrambled enough when EE saturates. 
Consequently, EE exhibits large fluctuation after saturation as shown in Fig.~\ref{figvelEEs} (I), 
because entanglement propagating from other parts of the system reaches the subsystem even after EE saturates in the vicinity of the subsystem.  
In the case of the CI model, $\xi\simeq 3$ for $d=2$ and thus $\xi\simeq d$. 

We briefly note that similar phenomena occur in the quench dynamics of spontaneous symmetry breaking.
Cooling the system through the transition temperature at a finite rate results in the formation of finite-size domains, in each of which the order parameter is chosen independently. 
The formation of domains has been observed in various condensed matter systems including Bose-Einstein condensates and superconductors \cite{delcampo14}. 
The mechanism for the formation of such domains, known as the Kibble-Zurek mechanism \cite{Kibble,Zurek}, 
was explained in terms of the finite propagation speed of light or causality in the context of cosmology \cite{Kibble}. 
Since causally disconnected regions do not influence each other, they have independent choices for the order parameter.
The formation of domains in the Kibble-Zurek mechanism is analogous to local saturation of EE in our problem. 
The independent choice of the order parameter in each domain leads to the loss of coherence in the Kibble-Zurek mechanism. 
This is analogous to the situation where the system does not thermalize when it is not scrambled enough in the CI model. 
Our result shows that the local thermalization time $t^\ast$s are almost the same for three models, though it is found that the ECI model achieves the thermalization. 
The significant difference between ECI model and the rests is the propagation speed of EE.
These facts can be understood if the Kibble-Zurek like mechanism takes places for the thermalization as discussed above. 
This interpretation is also consistent with the fact that the TI model does not thermalize irrespectively of the magnitude of the magnetic field, 
since the propagation speed of EE cannot exceed the maximum quasi-particle velocity. 
\par
Figure~\ref{TEBDCI} clearly shows that $t_{\mathrm{diff}}$ and $t^\ast$ hardly depend on the size of the system and $t_{\mathrm{diff}}> t^\ast$ up to $N=30$. We have also confirmed that thermalization is not achieved up to $N=14$ (Fig.~\ref{figSzTMLZ} III).
Based on these results, we expect that the CI model does not thermalize even in sufficiently large systems, which is consistent with the claim of Ref.~\cite{banuls11}. 
However, whether the CI model thermalizes or not in sufficiently large systems is beyond the scope of the present work, so we leave a detailed study of this issue in the future.
\par
In the present work, we have investigated the three models to focus on the two aspects of the system; integrability and NNN interactions. We note that a thermalizing model with only NN interactions should give deeper insights on thermalization, since there is no a priori reason that the model yields a short $t_{\mathrm{diff}}$. More case studies are indeed necessary for giving a definitive conclusion on the conjecture.

The simple setup for measuring the propagation speed of EE that we proposed in this paper may be easily applied to cold atom experiments. 
We specifically propose to use the setup for the experiments done by Greiner's group \cite{islam15,kaufman16}, 
in which R\'enyi entropy of ultracold bosonic atoms in a 1D optical lattice has been measured. 
The dynamics of R\'enyi entropy starting from the Mott insulating state that has been already measured corresponds to the dynamics of EE starting from the initial state (a). 
For realizing the initial state (b), we first prepare atoms in the Mott insulating regime and isolate two atoms on the left end of the chain by a high potential barrier. 
After ramping down the barrier potential between the two sites, we let the two sites evolve in time so that two atoms are entangled. 
The propagation speed of entanglement can be estimated by comparing the dynamics of R\'enyi entropy for the whole system except the single site on the left end analogous to the present study. 
An extension of the present work to dynamics of atoms in optical lattices is in progress \cite{yoshii20}.

\section*{Acknowledgments}
RY is grateful to E.\,Iyoda, S.\,Nakajima, M.\,Nitta, E.\,Saito, H.\,Tasaki, and A.\,Yamaguchi for fruitful discussions. 
ST thanks N.\,Hatano, T.\,Ikeda, M.\,Ueda, and A.\,Yoshinaga for valuable comments. 
We wish to thank one of the referees for his/her helpful comments to improve our manuscript.
The work of RY is supported by JSPS Grant-in-Aid for Scientific Research (KAKENHI Grant No.~19K14616 and 20H01838). 
The work of ST is supported by JSPS Grant-in-Aid for Scientific Research (KAKENHI Grant No.~19K03691).
ST was also supported by Chuo University Grant for Special Research.

\begin{appendix}

\section{Finite size effect}
\begin{figure}[t]
\includegraphics[width=7.5cm]{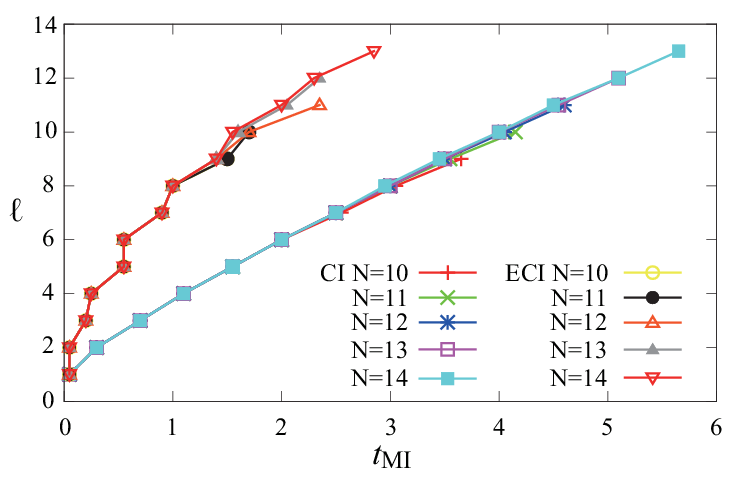}
\caption{Subsystem size dependence of $t_{\rm MI}$ for the CI and ECI models for different system sizes. } 
\label{figSizeDep}
\end{figure}

In this Appendix, we show that the finite-size effect is irrelevant to the evaluation of propagation of entanglement. We plot $t_{\rm MI}$ for different system sizes in Fig.\,\ref{figSizeDep} for the CI and ECI models. It clearly justifies the evaluation of the propagation speed of entanglement for $N=14$, because the finite size effect only appears in a few sites near the edge of the system, which does not affect the evaluation seriously.

\section{$t_{\rm diff}$ and $t_{\rm MI}$ for a larger threshold}\label{LargeThreshold}
\begin{figure}[t]
\includegraphics[width=7.5cm]{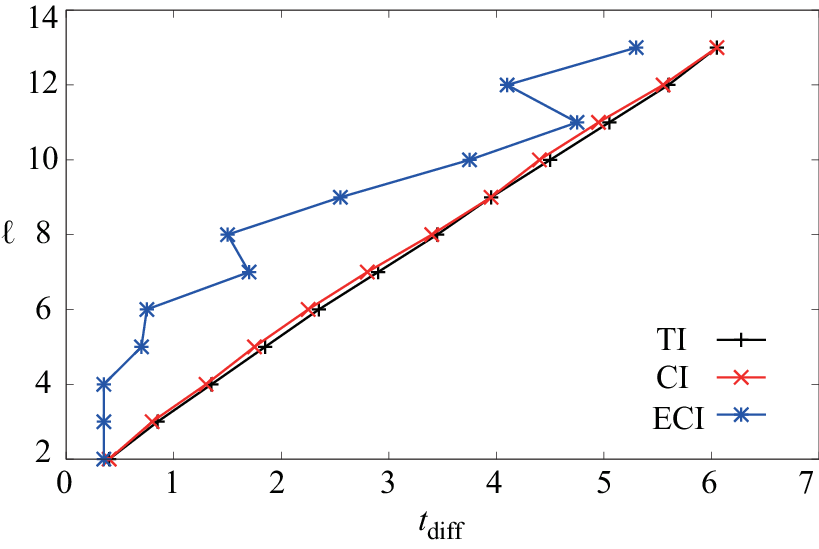}
\caption{Subsystem size dependence of $t_{\mathrm{diff}}$ for the TI model with $h=1.0$, CI model, and ECI models. 
Here we choose the thresholds to be $10^{-3}$.} 
\label{figEEthreshold-3}
\end{figure}

\begin{figure}[t]
\includegraphics[width=7.5cm]{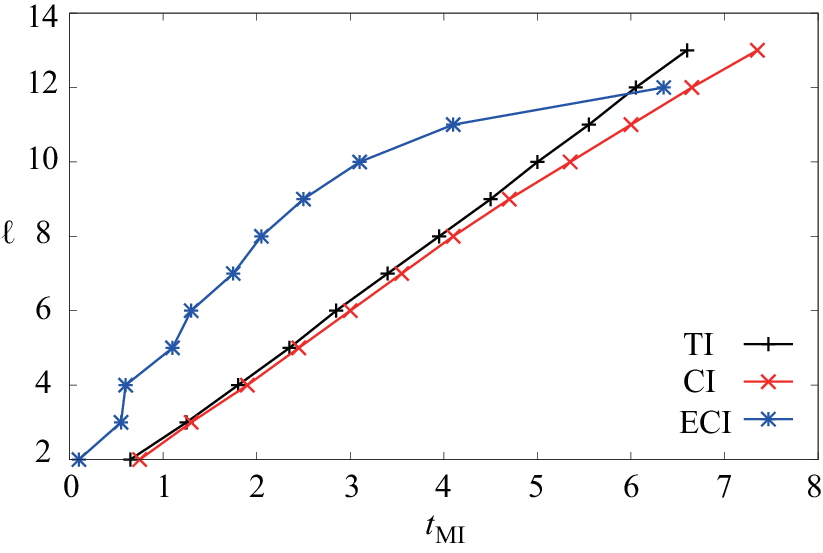}
\caption{Subsystem size dependence of $t_{\mathrm{MI}}$ for the TI model with $h=1.0$, CI model, and ECI models. 
Here we choose the thresholds to be $10^{-3}$.} 
\label{figMIthreshold-3}
\end{figure}

In our calculation, we have chosen the threshold $10^{-7}$ so that the velocities evaluated by $t_{\mathrm{diff}}$ and $t_{\mathrm{MI}}$ do not change in accuracy of 1 \% even if we choose smaller thresholds such as $10^{-8}$ or $10^{-9}$. However, for the experimental realizability, the threshold is too small. 
In fact, the propagation speed of entanglement can be estimated with lower accuracy even if we choose larger thresholds. Figures \ref{figEEthreshold-3} and \ref{figMIthreshold-3} show the $l$-dependence of $t_{\mathrm{diff}}$ and $t_{\mathrm{MI}}$ evaluated with the threshold $10^{-3}$, respectively. They demonstrate that the propagation speed of entanglement can be still estimated with lower accuracy. We thus expect that our proposal is still valid for experiments.

\section{Comparison of $t_{\mathrm{diff}}$ and $t_{\mathrm{MI}}$ with the time scale in Ref.\,\cite{iyoda18}}

In Ref. \cite{iyoda18}, Iyoda and Sagawa calculated the time evolution of mutual information and propose a time-scale that characterizes the propagation of entanglement, 
which we refer to $t_{\rm dump}$. 
In this Appendix, we give a comparison of $t_{\rm MI}$ and $t_{\rm diff}$ with $t_{\rm dump}$. $I(0|L)$ decreases from 2 at the initial moment. 
They define $t_{\rm dump}$ as the time at which $I(0|L)$ becomes less than 1.9 for the first time.

To make a comparison, we need to express $I(0|L)$ in terms of $I(0|R)$. Since the attached spin does not evolve in time, the EE for the $L+R$ system is a constant due to $S_0=S_{L+R}$. 
Here, $S_0=1$ for the initial state (7). Using the relations $S_{0+L}=S_R$ and $S_{L}=S_{0+R}$, one finds 
\begin{align}
I(0|R)&=S_0+S_R-S_{0+R}=S_0+S_{0+L}-S_{L}\nonumber\\
&=2S_0-(S_L+S_0-S_{0+L})=2-I(0|L).
\end{align}
Using Eq. (B1), $I(0|R)$ gets bigger than 0.1 at $t_{\rm dump}$. We plot $t_{\rm dump}$ as a function of the subsystem size $l$. 
One sees that the propagation speed estimated by $t_{\rm dump}$ is slightly less than that of quasiparticles.

\begin{figure}[t]
\includegraphics[width=7.5cm]{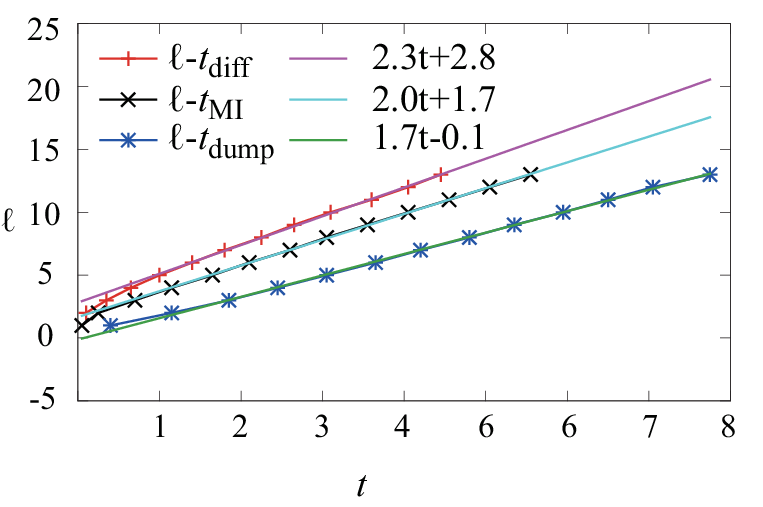}
\caption{Subsystem size dependence of $t_{\mathrm{diff}}$, $t_{\rm MI}$, and $t_{\mathrm{dump}}$. } 
\label{Fig_IyodaSagawa}
\end{figure}

\section{Comparison with the results obtained by time-evolving block decimation algorithm} 
\begin{figure}[t]
\includegraphics[width=7.5cm]{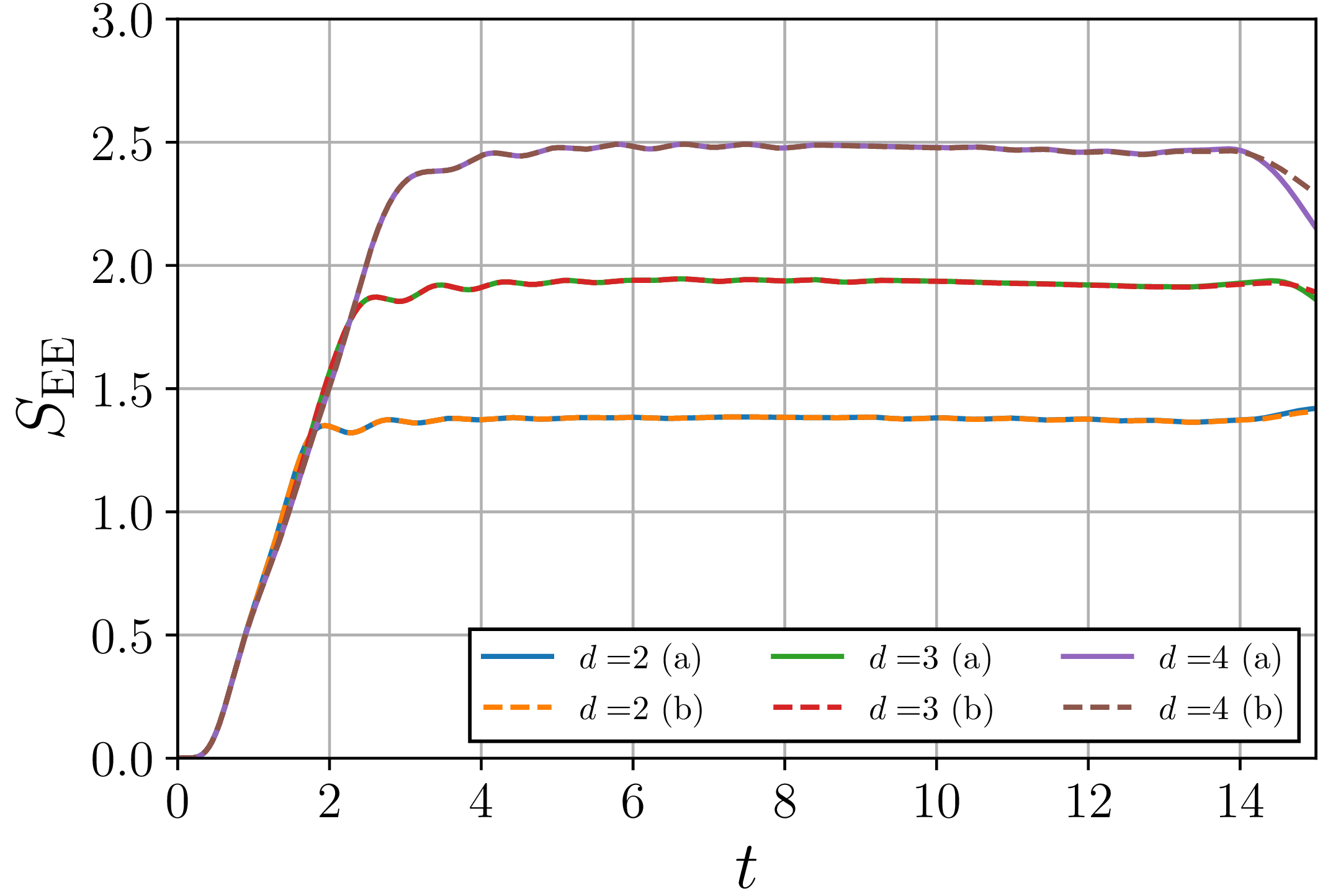}
\caption{Time-evolution of EE for TI model with $N=30$ obtained by the TEBD algorithm. Solid and dashed lines correspond to the initial states (a) and (b), respectively.} 
\label{TEBDTI}
\end{figure}

\begin{figure}
\includegraphics[width=7.5cm]{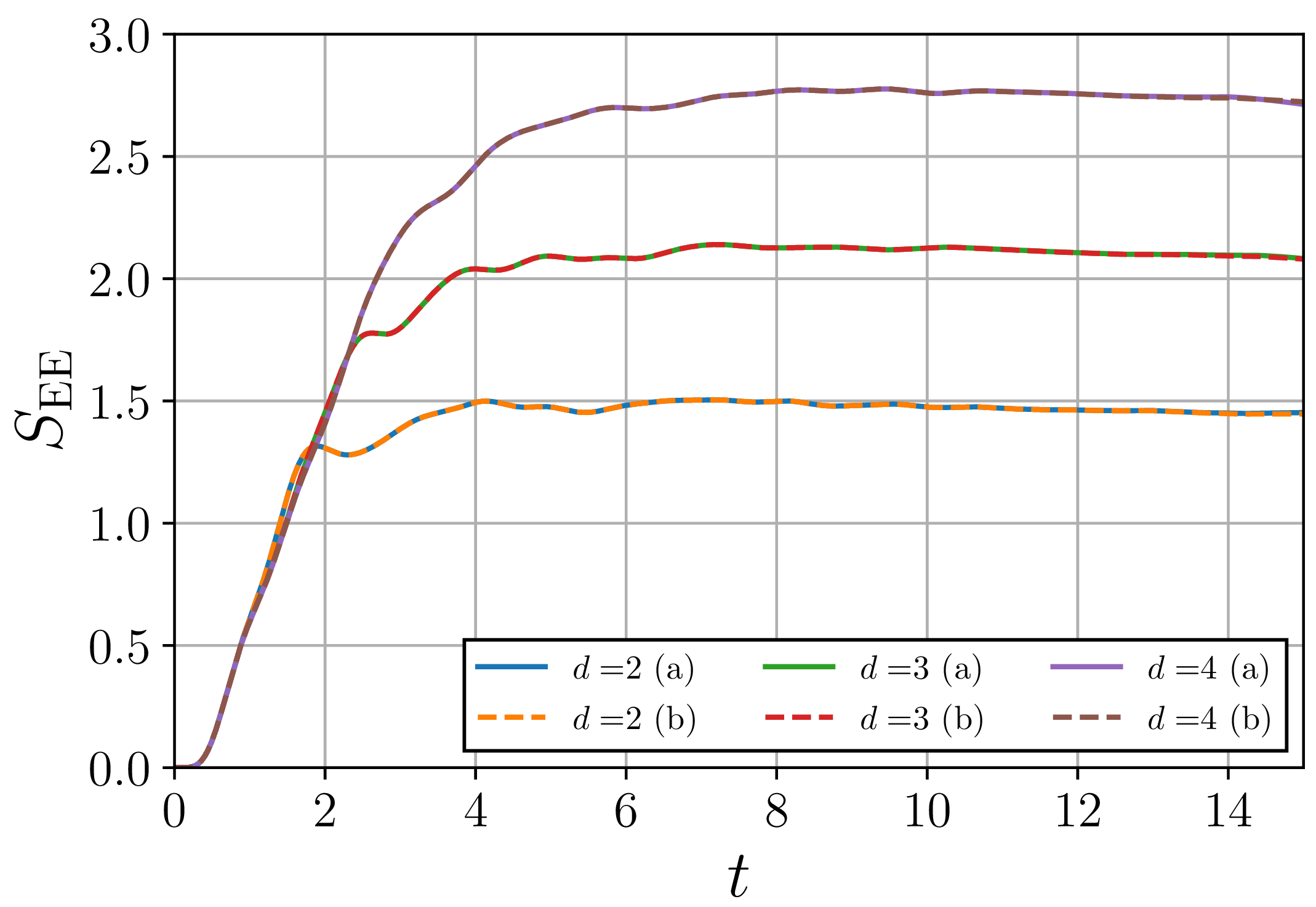}
\caption{Time-evolution of EE for CI model with $N=30$ obtained by the TEBD algorithm. Solid and dashed lines correspond to the initial states (a) and (b), respectively.} 
\label{TEBDCI}
\end{figure}

In this Appendix, we show the results obtained by the time-evolving block decimation (TEBD) algorithm \cite{TEBD}. 
This method enables one to analyze the larger size systems at the expense of the numerical accuracy. 
We employ the 2nd-order Suzuki-Trotter decomposition with time step $dt = 10^{-2}$.
The time-evolution of EE for TI and CI models with $N=30$ obtained by TEBD are shown in Figs.\,\ref{TEBDTI} and \ref{TEBDCI}, respectively. 
The solid and the dashed lines correspond to, respectively, the different initial states (a) and (b) (see Fig.\ref{figmodel} (a) and (b)). 
Both figures show the smaller fluctuations compared to the results for $N=14$ in Fig.~3 due to the suppression of the finite size effect. 
 

\begin{figure}
\includegraphics[width=7.5cm]{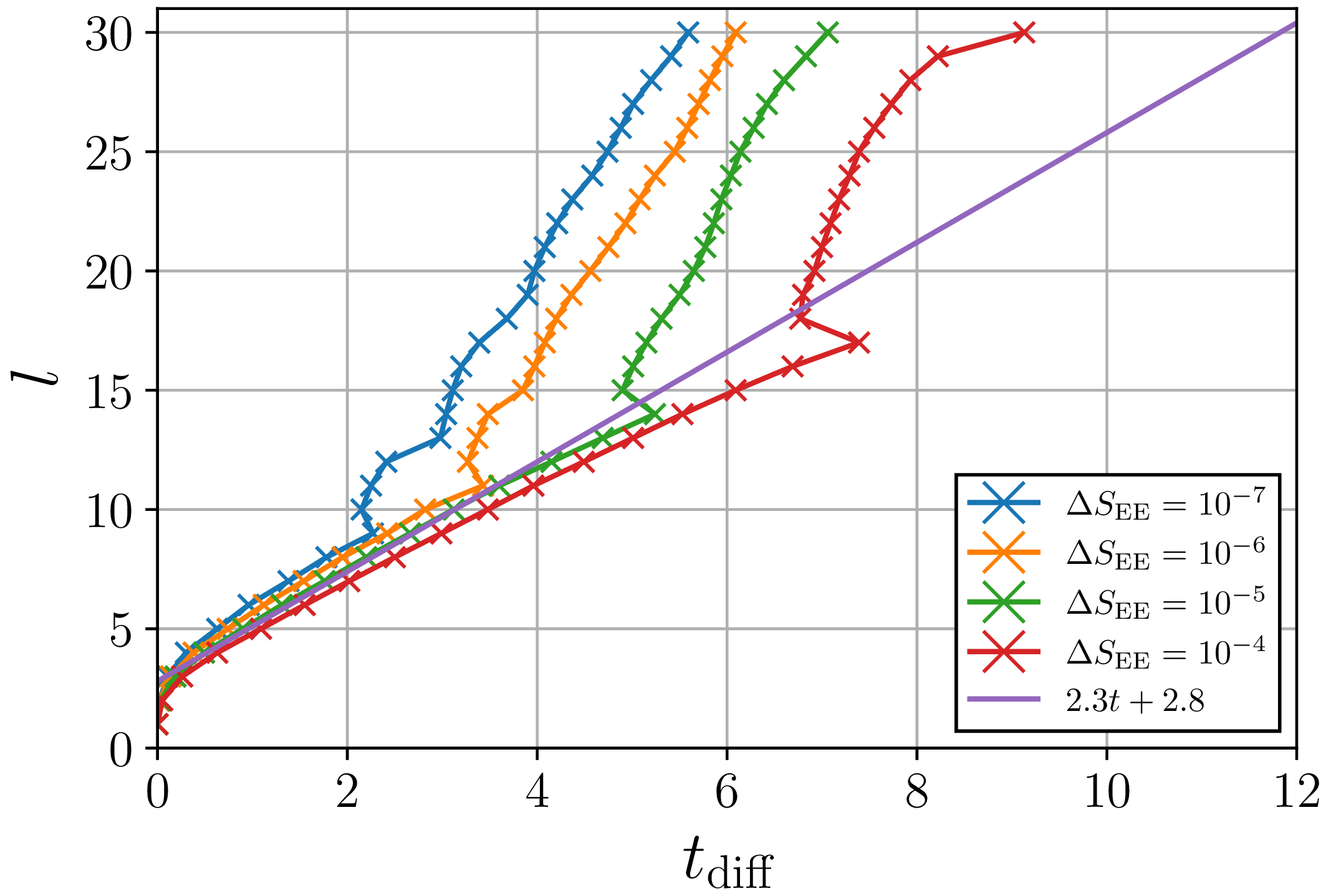}
\caption{Propagation speed of EE  evaluated by the TEBD algorithm for TI model with $N=30$ for various thresholds $\Delta S_{\mathrm{EE}}$.} 
\label{TEBDVTI}
\end{figure}

\begin{figure}
\includegraphics[width=7.5cm]{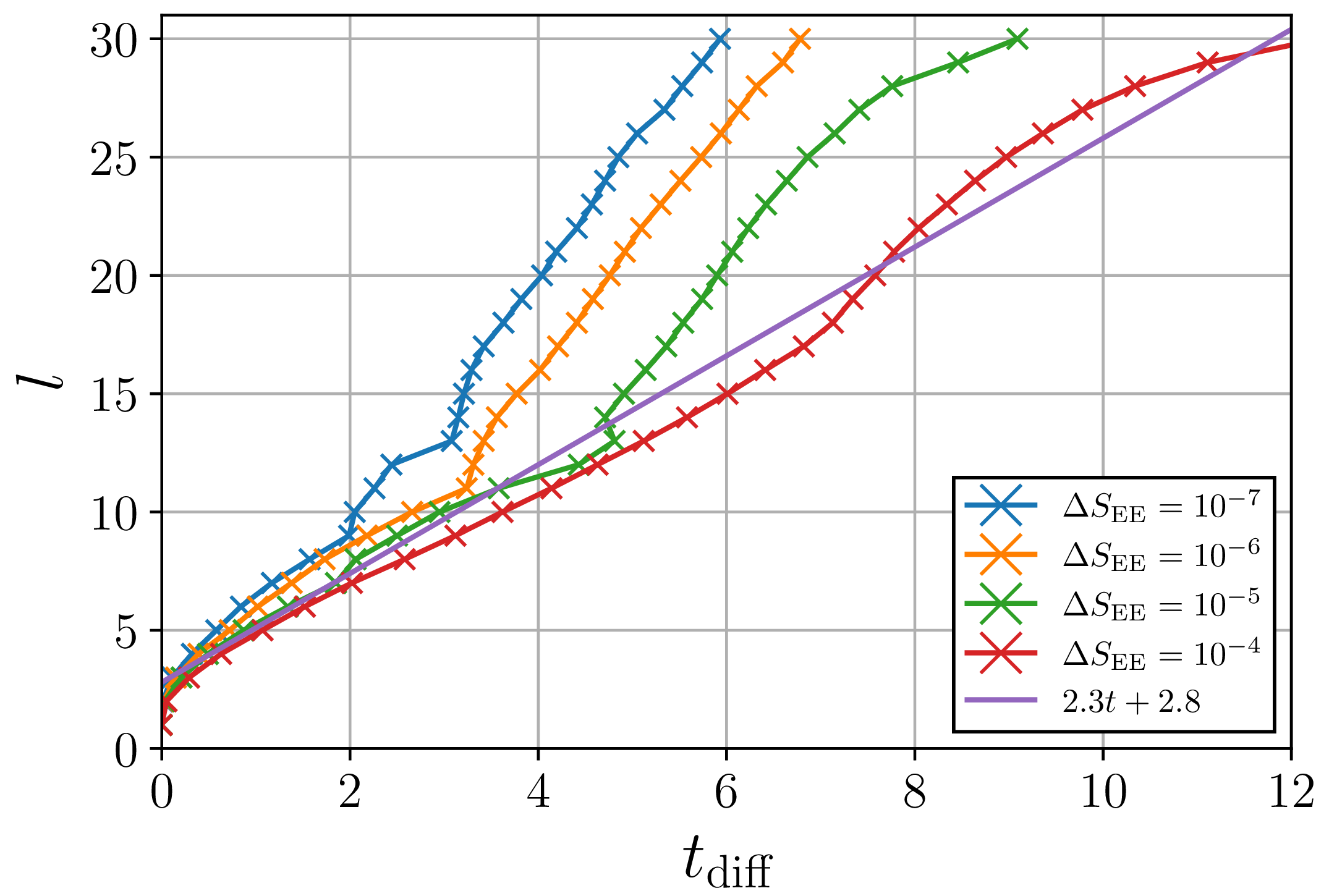}
\caption{Propagation speed of EE  evaluated by the TEBD algorithm for CI model with $N=30$ for various thresholds $\Delta S_{\mathrm{EE}}$.}
\label{TEBDVCI}
\end{figure}

Figures~\ref{TEBDVTI} and \ref{TEBDVCI} show the propagation speed of EE for TI and CI models, respectively, evaluated by the TEBD algorithm with various thresholds $\Delta S_{\mathrm{EE}}$. 
These figures clearly show that the TEBD algorithm does not provide with enough numerical accuracy in our setup. 
To evaluate accurately the propagation speed of EE, it is required to employ the higher order Suzuki-Trotter decomposition and/or smaller time step.

\end{appendix}

\end{document}